\documentclass[letterpaper]{article} 
\usepackage[preprint]{aaai2027}  
\usepackage[hyphens]{url}  
\usepackage{graphicx} 
\usepackage{natbib}  
\usepackage{caption} 
\usepackage{algorithm}
\usepackage{algorithmic}

\usepackage{newfloat}
\usepackage{listings}
\DeclareCaptionStyle{ruled}{labelfont=normalfont,labelsep=colon,strut=off} 
\floatstyle{ruled}
\newfloat{listing}{tb}{lst}{}
\floatname{listing}{Listing}

\usepackage{booktabs}

\title{Beyond Evidence Coverage: Encoding-Time Access in Multi-Agent Systems}

\title{Beyond Evidence Coverage: Encoding-Time Access in Multi-Agent Systems}

\author{
Joshua Huang\equalcontrib,
Ajit Mallavarapu\equalcontrib,
Elliott Kalt
}

\affiliations{
Cornell Tech\\
New York, NY, USA\\
\{jh3262, am3574, esk226\} @cornell.edu
}

\usepackage{amsmath}

\newcommand{\blocked}{\textsc{Blocked}}

\newcommand{\matched}{\textsc{Matched}}

\newcommand{\gold}{\textsc{Gold}}

\newcommand{\dist}{\textsc{Distractor}}

\newcommand{\nogold}{\textsc{NoGold}}

\newcommand{\full}{\textsc{Full}}

\begin{document}

\maketitle

\begin{abstract}
Cache-sharing multi-agent systems can retain every evidence token's representation while preventing cross-worker access during encoding. We test whether this coverage recovers joint-encoding quality using attention masks over two logical workers in frozen Qwen3-4B, Qwen3-8B, and Llama-3.1-8B-Instruct models. Tokens, positions, and answer-time access remain fixed within comparisons. We evaluate 200 two-hop MuSiQue-Ans questions and 200 selected HotpotQA bridge questions. In data-layout seed 0, averaged across models, blocking access costs 5.6 and 6.8 F1 points relative to full access on MuSiQue and HotpotQA, respectively. Exposing the earlier worker's supporting paragraph improves F1 over blocked access by 7.6 and 6.7 points; equal-token non-support access provides no detectable average gain. With model weights and greedy decoding fixed, we vary distractor layouts using five data-layout seeds per dataset for Qwen3-8B and three for Llama-3.1-8B-Instruct. Supporting-paragraph access has higher mean F1 than blocked and equal-token non-support access in every evaluated data-layout seed. Supporting access can also outperform full access, but its advantage depends on the model, dataset, and layout. Oracle support labels isolate access from selection error; the study does not evaluate a deployable selector. Complete evidence coverage at answer time therefore does not guarantee joint-encoding quality: which supporting facts can interact during encoding matters.

\end{abstract}


\section{Introduction}

Multi-agent language-model systems distribute work across models that exchange messages or internal representations \citep{wu2023autogen,zou2025latentmas}. For evidence gathering, a natural design assigns documents to parallel workers and combines their outputs at an orchestrator. Passing each worker's full key--value (KV) cache lets the answering model access a representation of every evidence token without requiring the worker to select which facts to include in a written summary. LatentMAS formalizes information-preserving transfer of cached working memory \citep{zou2025latentmas}. Such transfer preserves the states already computed; it does not establish that merging independently constructed caches reproduces a jointly encoded context. Each token's representation depends on the evidence its worker could access during encoding.

Consider answering who governs a person's favorite city. One worker receives the person-to-city fact; another receives the city-to-mayor fact. Their caches can preserve representations of both facts even though neither worker encoded them together. We ask: \emph{what does blocking cross-worker access during encoding cost, and can access to just the supporting evidence recover that cost when the answerer already has access to every evidence position?}

We investigate this question through controlled attention masks over two logical workers in frozen Qwen3-4B and Qwen3-8B \citep{qwen3technicalreport}, and Llama-3.1-8B-Instruct \citep{grattafiori2024llama3herdmodels} (hereafter Llama-8B). Within each layout, we keep the question, tokens, positions, and decoding protocol fixed, and vary which earlier-block positions the later block can attend to during encoding. The answerer retains access to every evidence position. This isolates the encoding-access intervention from omitted report content and changes to document order. We evaluate 200 two-hop MuSiQue-Ans questions \citep{trivedi2022musique} (hereafter MuSiQue) and 200 selected HotpotQA bridge questions \citep{yang-etal-2018-hotpotqa}, comparing supporting-paragraph access with blocked, full, and equal-token non-support access. Oracle support labels separate the effect of evidence access from errors in selecting what to expose.

Across both datasets, supporting-paragraph access improves performance over blocked access and the equal-token control. These gains persist across randomized distractor layouts in both 8B models, with model weights and greedy decoding held fixed. Selective access also outperforms full access in some settings, although this secondary finding depends on the model, dataset, and evidence arrangement. MuSiQue follow-ups and implementation checks further examine the scope of these effects. These findings complement system-level failure taxonomies \citep{cemri2025fail} with a controlled account of evidence sharing: retaining every evidence token's cached representation at answer time does not guarantee joint-encoding quality. Which supporting facts can interact while those representations are constructed matters.

\section{Related Work}

\paragraph{Communication in multi-agent language-model systems.} 
Multi-agent frameworks organize collaboration through text exchanged among role-specialized agents \citep{wu2023autogen}; recent systems instead pass internal state, transferring latent working memory or KV caches between models \citep{zou2025latentmas}. Analyses catalog coordination and verification failures \citep{cemri2025fail}, test specialist KV payloads under adversarial modification \citep{brito2026integrity}, and audit whether relayed caches carry example-specific information \citep{cheng2026latent}. CanonicalMerge addresses input-order dependence by imposing a deterministic, content-based ordering on cached fragments \citep{baquero2026cachemergingconvergentreplicated}. Its guarantees concern the assembly of already computed states. Our study examines which evidence is accessible during state construction, while retaining every evidence position for answering.

\paragraph{Independent passage encoding and evidence fusion.} 

Models trained to encode passages independently and combine their representations during decoding can aggregate evidence effectively \citep{izacard2021fid}. Parallel Context Windows (PCW) reports gains on comparison questions but losses on bridge questions in off-the-shelf models, attributing the latter to missing conditioning between evidence texts \citep{ratner2023pcw}. Graph-KV enables target segments to attend to selected source-segment caches through a graph-structured attention mask, improving bridge and compositional question answering over independent encoding \citep{wang2025graphkv}. Both PCW and Graph-KV also modify positional assignments. Our comparison holds tokens, positions, and answer-time access fixed and tests supporting-paragraph access against equally many non-support tokens. At the level of evidence availability, \gold{} can be viewed as an oracle pattern of cross-block connections. Whether retrieved or learned connections can reproduce its benefits without support labels remains an open question in our setting.

\paragraph{KV-cache reuse and cross-document dependencies.} 
CacheBlend selectively recomputes token representations to recover interactions missing from independently cached chunks \citep{yao2024cacheblend}. ProphetKV uses query-guided token selection and a two-stage recomputation pipeline \citep{wang2026prophetkv}, while InfoFlow KV combines attention-based selection with global positional assignments and chunk reordering \citep{teng2026infoflow}. RelaxKV jointly selects repair targets and their attention context while preserving global positions, including controls that fix repair targets and vary recomputation context \citep{qi2026relaxkvrecomputationguidedquery}. CacheWeave adapts the backbone to independently cached chunks and trains a block selector using model-derived oracle supervision \citep{pan2026cacheweavelearningrecomputeefficient}. CoinRAG composes contextualized KV slices from parent chunks, with positional alignment and nugget-aware fine-tuning; its comparison with isolated nugget encoding also highlights the importance of construction context \citep{kim2026coinragcontextualizedinformationnugget}. Related analyses examine distractor structure and compatibility between reused and recomputed states \citep{peng2026kvcacherepairbudgetgo}, and changes to judge decisions when reuse weakens cross-candidate interactions \citep{liang2026reuse}. These studies establish the importance of cross-document context and selective recomputation. Our diagnostic uses annotated supporting paragraphs to compare support access with blocked, full, and equal-token non-support access, removing selector error from that comparison. Each mask condition recomputes the evidence representations, so the experiment does not establish the efficiency of a cache-repair policy.

\paragraph{Context position, document order, and attention.} Evidence availability alone does not ensure effective use. \citet{liu2023middle} show that language model performance depends strongly on where relevant information appears within a long context. For multi-hop questions, \citet{yu2024repetition} identify sensitivity to the relative order of supporting documents and propose context repetition to improve reasoning over misordered evidence. \citet{huang2025masking} examine document permutations and modified attention masks, connecting multi-hop performance to the interactions permitted by the model architecture. Within each layout, our mask comparisons preserve the serialized input and change only the later block's access to earlier evidence during encoding. Separate follow-ups move supporting paragraphs or place both supports in the same block to test how the access effect depends on evidence arrangement.

\paragraph{Evaluating evidence composition.} A question's compositional wording does not guarantee that answering requires multiple supporting paragraphs. \citet{min2019composition} demonstrate substantial single-hop shortcuts in a multi-hop benchmark. MuSiQue addresses this issue by composing connected single-hop questions and applying filters intended to reduce disconnected reasoning \citep{trivedi2022musique}. Synthetic studies also expose limitations that aggregate benchmark scores can conceal: \citet{guo2025twohop} find strong distractor sensitivity in pretrained models and analyze how a sequential retrieval mechanism emerges in trained transformers. These studies motivate our combination of synthetic chains, two-hop MuSiQue, HotpotQA bridge questions \citep{yang-etal-2018-hotpotqa}, and single-support probes. Our MuSiQue and HotpotQA cohorts are fixed independently of model correctness; probe outcomes define screening sensitivity subsets. The behavioral interventions do not identify the internal retrieval mechanism used by the evaluated models.

\paragraph{Answer-time reasoning and retrieval.} Chain-of-thought prompting can improve performance on complex reasoning tasks by eliciting intermediate reasoning steps \citep{wei2022cot}. IRCoT interleaves reasoning with retrieval, allowing intermediate conclusions to guide subsequent evidence acquisition \citep{trivedi2023ircot}. Both operate during answering; IRCoT also changes which evidence is supplied. Our supplementary experiments on smaller MuSiQue cohorts and synthetic chains keep the supplied evidence fixed and examine whether answer-time reasoning compensates for restricted encoding access under the tested generation budgets.

\section{Controlled Encoding}

Let $S$ be a shared system-and-question prefix and $A$,$B$ be evidence blocks, with $A$ preceding $B$. The workers are logical readers within one frozen transformer. During evidence encoding, each block retains causal attention within itself and access to $S$. The answer prompt and generated tokens can attend to every prefix and evidence position in all conditions. Figure~\ref{fig:design} in Appendix~\ref{app:setup} illustrates the encoding-access intervention.

We recompute the same sequence $[S,A,B]$ with an additive attention mask. Let $V$ contain the positions in $A$ that are accessible to $B$. For a query position $i$ in $B$ and a key position $j$ in $A$,
\begin{equation}
M^{(V)}_{ij} =
\begin{cases}
0, & j \in V,\\
-\infty, & j \notin V.
\end{cases}
\label{eq:mask}
\end{equation}
All other causal edges remain unchanged. In implementation, blocked entries use the minimum finite value of the model dtype.

The same mask is applied to every attention head at every layer, enforcing block-level availability throughout the network. This intervention does not identify which heads or layers are responsible for the effect, or whether restricting the intervention to a subset would suffice.

We compare four masks: \blocked{} exposes no positions in $A$; \gold{} exposes its supporting paragraph; \matched{} exposes the same number of tokens from non-supporting paragraphs; and \full{} exposes all of $A$, including separators. Paragraph spans are identified from offsets in the unchanged block tokenization. In the primary layouts, matched tokens precede $A$'s support and come from paragraphs without recognized answer-alias or bridge mentions. Matching token counts does not match positions or all information carried by the selected contextual states.

Every condition retains all supplied paragraphs: 20 for MuSiQue and 10 for HotpotQA. Oracle support labels determine which attention edges open; no label enters the prompt. The comparison therefore measures access under oracle selection, without evaluating a practical evidence selector.

We also report native joint encoding ($B1$) and stitched caches ($K0$). $B1$ uses standard causal encoding of the serialized input. $K0$ independently encodes each block at local positions, shifts rotary keys to their assembled positions, and concatenates the caches. $B1$--$K0$ therefore changes both encoding access and cache-position handling. Our primary effects compare masks within one implementation; we do not assume numerical equivalence between \blocked{} and separately encoded blocks at the same positions.

\section{Experimental Setup}

\paragraph{MuSiQue.}
The cohort contains 200 previously unused two-hop MuSiQue-Ans development questions, selected in fixed benchmark order after grounding and tokenizer-feasibility checks. All 20 supplied paragraphs are retained, with 10 per worker. $AB$ places the hop-1 support in $A$ and the hop-2 support in $B$; $BA$ reverses their assignment. Each support is last in its block.

\paragraph{HotpotQA.}
The cohort contains 200 bridge questions from the distractor development set \citep{yang-etal-2018-hotpotqa}, selected in fixed order without using model outputs. HotpotQA labels neither hop order nor bridge entities. We designate the uniquely answer-containing support as hop 2 and infer a bridge from a supporting title mentioned in the other support; questions failing either requirement are excluded (Appendix~\ref{app:setup}). All 10 supplied paragraphs are retained, with five per worker, in $AB$ and $BA$ layouts with each support last in its block.

\paragraph{Follow-ups and probes.}
On MuSiQue at data-layout seed 0, both 8B models additionally move $A$'s support to its first position or place both supports together in $A$ ($AA$) or $B$ ($BB$). Closed-book and single-support probes define sensitivity subsets without changing either primary cohort. Synthetic and earlier reasoning experiments appear in Appendices~\ref{app:tables} and \ref{app:thinking4b}.  

\paragraph{Models and decoding.} 
All models are frozen and use greedy decoding with a 1,024-token answer cap. The Qwen3 models use an empty thinking prefill; Llama-8B uses its native assistant prefix. Prompts request a short boxed answer. For each dataset, we evaluate the four masks with five data-layout seeds for Qwen3-8B, three for Llama-8B, and one for Qwen3-4B, including seed 0 for every model. These seeds vary distractor assignment and ordering within fixed question cohorts. $B1$ and $K0$ results are reported at data-layout seed 0. Runtime and probe details appear in Appendix~\ref{app:setup}; supplementary forced-thinking protocols appear in Appendices~\ref{app:tables} and \ref{app:thinking4b}. Forced thinking is not evaluated on either primary cohort.

\paragraph{Scoring and uncertainty.} We extract the boxed answer, with fallbacks when absent, after removing delimited thinking. We report normalized exact match (EM) and token $F1$, maximizing over reference aliases; HotpotQA supplies one reference answer per question. Normalization lowercases, removes articles and punctuation, collapses whitespace, and maps recognized number words to digits. Capped answers remain in the scores.

We average paired differences over $AB$/$BA$, or $AA$/$BB$, within each source. Intervals are descriptive 95\% percentile intervals from 20,000 source-bootstrap resamples with seed 0. Exploratory pooled estimates also average models within source before resampling; their intervals condition on the evaluated models. Comparisons are unadjusted for multiplicity. These intervals quantify variation across source questions for the fixed layouts, selected windows, and models; they do not estimate variability across layout or control-selection seeds. 

\section{Results and Analysis}

\paragraph{Supporting-paragraph access improves performance on both datasets.} Averaged across the three models, \gold{} improves over \blocked{} by 7.6 F1 points $[5.1,10.2]$ on MuSiQue and 6.7 $[4.3,9.2]$ on HotpotQA. The corresponding \full{}--\blocked{} differences are 5.6 $[2.8,8.5]$ and 6.8 $[4.3,9.2]$. Supporting-paragraph access therefore recovers the observed loss in pooled mean performance on MuSiQue and approximately matches that loss on HotpotQA.

The equal-token control provides no detectable pooled gain over \blocked{}: $-0.8$ F1 $[-2.6,1.0]$ on MuSiQue and $-0.3$ $[-1.9,1.4]$ on HotpotQA. \gold{} exceeds this control by 8.4 $[5.7,11.3]$ and 7.0 $[4.6,9.5]$ points, respectively. Both \gold{}--\blocked{} and \gold{}--\matched{} have positive intervals in every model on both datasets (Table~\ref{tab:effects}). The gain thus depends on which tokens are exposed, beyond their number.

\begin{table}[t]
\centering
\footnotesize
\begin{tabular}{@{}lccc@{}}
\toprule
Contrast & Qwen3-4B & Qwen3-8B & Llama-8B\\
\midrule
\multicolumn{4}{@{}l}{MuSiQue}\\
\gold{}--\blocked{} & $+9.8$ & $+7.4$ & $+5.6$\\
 & $[6.5,13.2]$ & $[3.5,11.5]$ & $[1.8,9.5]$\\
\gold{}--\matched{} & $+11.3$ & $+7.0$ & $+7.0$\\
 & $[7.4,15.2]$ & $[3.1,11.1]$ & $[3.1,11.0]$\\
\gold{}--\full{} & $+3.6$ & $+1.0$ & $+1.3$\\
 & $[0.6,6.7]$ & $[-2.2,4.1]$ & $[-2.4,5.2]$\\
\full{}--\blocked{} & $+6.2$ & $+6.4$ & $+4.3$\\
 & $[2.5,9.8]$ & $[2.3,10.7]$ & $[-0.1,8.6]$\\
\midrule
\multicolumn{4}{@{}l}{HotpotQA}\\
\gold{}--\blocked{} & $+8.7$ & $+6.6$ & $+4.9$\\
 & $[5.3,12.3]$ & $[3.1,10.2]$ & $[1.1,8.8]$\\
\gold{}--\matched{} & $+8.7$ & $+5.7$ & $+6.6$\\
 & $[5.3,12.3]$ & $[2.1,9.3]$ & $[2.8,10.5]$\\
\gold{}--\full{} & $+0.5$ & $+1.4$ & $-1.9$\\
 & $[-1.6,2.6]$ & $[-0.4,3.3]$ & $[-4.7,0.7]$\\
\full{}--\blocked{} & $+8.2$ & $+5.2$ & $+6.9$\\
 & $[4.6,12.0]$ & $[1.6,8.9]$ & $[3.2,10.5]$\\
\bottomrule
\end{tabular}
\caption{Paired F1 differences and 95\% source-bootstrap intervals at data-layout seed 0, using 200 questions per dataset and model. An interval crossing zero does not establish equivalence.}
\label{tab:effects}
\end{table}

\paragraph{Support-access gains persist across distractor layouts.} Across five data-layout seeds for Qwen3-8B and three for Llama-8B on each dataset, \gold{} has higher mean $F1$ than both \blocked{} and \matched{} in every seed. \full{} also exceeds \blocked{} throughout. Table~\ref{tab:multiseed} (Appendix~\ref{app:tables}) summarizes the mean differences. 

\paragraph{Selective access has a conditional advantage.} At data-layout seed 0, \gold{} exceeds \full{} by 3.6 F1 points $[0.6,6.7]$ for Qwen3-4B on MuSiQue; the corresponding intervals cross zero in the other five model--dataset comparisons (Table~\ref{tab:effects}). Across data-layout seeds, \gold{} exceeds \full{} in four of five Qwen3-8B runs on each dataset and in two of three Llama-8B runs on MuSiQue, with mean differences under one point (Table~\ref{tab:multiseed}, Appendix~\ref{app:tables}). On HotpotQA, \full{} instead scores higher in all three Llama-8B runs, by 1.1 F1 points on average.

\paragraph{Co-located supports show no detectable average benefit.}
In the MuSiQue follow-up at data-layout seed 0, pooled across the two 8B models, \full{} improves over \blocked{} by 5.4 F1 points $[2.0,8.8]$ when supports are split, but by $-0.8$ $[-3.3,1.7]$ when both are in one worker. The paired difference between these access effects is 6.2 points $[2.5,9.9]$, with positive intervals in each model. Co-location also changes support adjacency, hop order, and distractor allocation, so this interaction does not isolate the worker boundary alone.

\paragraph{Support specificity persists across positions.}
Moving $A$'s support from last to first in the MuSiQue seed-0 layouts, while placing the equal-token control near $B$, preserves \gold{}--\matched{} gains: 9.9 F1 $[5.7,14.2]$ for Qwen3-8B and 7.9 $[3.9,12.0]$ for Llama-8B. Proximity to $B$ alone therefore does not explain the support benefit. In the same layouts, \gold{}--\full{} changes from $+1.0$ to $-3.1$ for Qwen3-8B and from $+1.3$ to $-1.2$ for Llama-8B.

\paragraph{Screening sensitivity.}
On sources where the closed-book and both single-support probes score F1 below 0.5, \gold{}--\blocked{} and \gold{}--\matched{} point estimates remain positive for every model on both datasets, with two HotpotQA intervals including zero (Appendix~\ref{app:screening}).



\paragraph{Implementation checks and stitched-cache comparisons.}
On 48 diagnostic MuSiQue layouts per 8B model, replacing $A$'s content leaves $B$'s cached states bit-identical at every layer under \blocked{}. With a common attention kernel, \full{} and $B1$ caches and first-answer logits also match exactly on these layouts. The blocked mask remains numerically nonidentical to separate encoding at the same positions (Appendix~\ref{app:setup}). At data-layout seed 0, $B1$ exceeds $K0$ by 5.5 F1 points $[2.9,8.1]$ on MuSiQue and 6.1 $[3.7,8.6]$ on HotpotQA, pooled across models. The masks split this gap exactly, source by source, with positions held fixed. On MuSiQue, $B1$--\full{} is $+0.5$ $[-0.5,1.6]$, \full{}--\blocked{} is $+5.6$ $[2.8,8.5]$, and \blocked{}--$K0$ is $-0.7$ $[-2.5,1.2]$; on HotpotQA the three steps are $-0.7$ $[-1.3,-0.1]$, $+6.8$ $[4.3,9.2]$, and $+0.1$ $[-1.3,1.4]$. The access step carries the gap on both datasets.

\paragraph{Answer-time reasoning has mixed effects in supplementary experiments.}
With Qwen3-8B, forced thinking raises \blocked{} F1 from 21.7 to 47.6 on an earlier 25-source MuSiQue sample, against 57.2 for \full{}, but lowers \blocked{} accuracy from 89\% to 41\% on synthetic chains when a look-alike distractor is in the other worker. Forced thinking is not evaluated on either primary cohort (Appendix~\ref{app:tables}).



\section{Limitations and Conclusion}

The evaluation covers 200 two-hop MuSiQue questions and 200 HotpotQA bridge questions, three frozen models from two families, and two logical readers. Qwen3-8B and Llama-8B are evaluated over five and three data-layout seeds per dataset; Qwen3-4B, the MuSiQue follow-ups, and all pooled estimates use one. Larger models, more hops, and more workers remain untested. HotpotQA hop order and bridge entities are inferred by rule; cohort selection and evidence counts limit cross-dataset score comparisons. Support selection is oracle, and matching token counts does not match positions or causal prefixes. We evaluate neither fine-tuning for independent encoding \citep{ma2025blockattention} nor forced thinking on the primary cohorts. Full recomputation establishes no efficiency gain. 

In the evaluated frozen models, retaining every evidence token's representation at answer time does not guarantee joint-encoding quality. On both datasets, access to one supporting paragraph recovers the average loss from blocking cross-worker interaction, and equally many non-support tokens do not. Learning which evidence to expose without oracle labels is the next practical test.

\bibliography{aaai2027}

@misc{wu2023autogen,
  author = {Wu, Qingyun and Bansal, Gagan and Zhang, Jieyu
    and Wu, Yiran and Li, Beibin and Zhu, Erkang and Jiang, Li
    and Zhang, Xiaoyun and Zhang, Shaokun and Liu, Jiale
    and Awadallah, Ahmed Hassan and White, Ryen W.
    and Burger, Doug and Wang, Chi},
  title = {{AutoGen}: Enabling Next-Gen {LLM} Applications
    via Multi-Agent Conversation},
  year = {2023},
  howpublished = {arXiv:2308.08155},
  url = {https://arxiv.org/abs/2308.08155}
}

@misc{zou2025latentmas,
  author = {Zou, Jiaru and Qiu, Ruizhong and Li, Gaotang
    and Yang, Xiyuan and Tieu, Katherine and Lu, Pan
    and Shen, Ke and Tong, Hanghang and Choi, Yejin
    and He, Jingrui and Zou, James and Wang, Mengdi
    and Yang, Ling},
  title = {Latent Collaboration in Multi-Agent Systems},
  year = {2025},
  howpublished = {arXiv:2511.20639},
  url = {https://arxiv.org/abs/2511.20639}
}

@misc{cemri2025fail,
  author = {Cemri, Mert and Pan, Melissa Z. and Yang, Shuyi
    and Agrawal, Lakshya A. and Chopra, Bhavya and Tiwari, Rishabh
    and Keutzer, Kurt and Parameswaran, Aditya and Klein, Dan
    and Ramchandran, Kannan and Zaharia, Matei
    and Gonzalez, Joseph E. and Stoica, Ion},
  title = {Why Do Multi-Agent {LLM} Systems Fail?},
  year = {2025},
  howpublished = {arXiv:2503.13657},
  url = {https://arxiv.org/abs/2503.13657}
}

@misc{brito2026integrity,
  author = {Brito, Lu{\'i}s and Baquero, Carlos},
  title = {When Latent Agents Lie: {KV}-Cache Integrity
    in Multi-Agent {LLM} Collaboration},
  year = {2026},
  howpublished = {arXiv:2606.28958},
  url = {https://arxiv.org/abs/2606.28958}
}

@inproceedings{ratner2023pcw,
  author = {Ratner, Nir and Levine, Yoav and Belinkov, Yonatan
    and Ram, Ori and Magar, Inbal and Abend, Omri
    and Karpas, Ehud and Shashua, Amnon
    and Leyton-Brown, Kevin and Shoham, Yoav},
  title = {Parallel Context Windows for Large Language Models},
  booktitle = {Proceedings of ACL},
  pages = {6383--6402},
  year = {2023},
  doi = {10.18653/v1/2023.acl-long.352},
  url = {https://aclanthology.org/2023.acl-long.352/}
}

@inproceedings{huang2025masking,
  author = {Huang, Wenyu and Vougiouklis, Pavlos
    and Lapata, Mirella and Pan, Jeff Z.},
  title = {Masking in Multi-hop {QA}: An Analysis of How
    Language Models Perform with Context Permutation},
  booktitle = {Proceedings of ACL},
  pages = {17781--17795},
  year = {2025},
  doi = {10.18653/v1/2025.acl-long.869},
  url = {https://aclanthology.org/2025.acl-long.869/}
}

@misc{liang2026reuse,
  author = {Liang, Sichu and Wang, Zhenglin and Chu, Jiajia
    and Xia, Pengfei and Zang, Hui and Zhou, Deyu},
  title = {When {KV} Cache Reuse Fails in Multi-Agent Systems:
    Cross-Candidate Interaction is Crucial for {LLM} Judges},
  year = {2026},
  howpublished = {arXiv:2601.08343},
  url = {https://arxiv.org/abs/2601.08343}
}

@misc{yao2024cacheblend,
  author = {Yao, Jiayi and Li, Hanchen and Liu, Yuhan
    and Ray, Siddhant and Cheng, Yihua and Zhang, Qizheng
    and Du, Kuntai and Lu, Shan and Jiang, Junchen},
  title = {{CacheBlend}: Fast Large Language Model Serving
    for {RAG} with Cached Knowledge Fusion},
  year = {2024},
  howpublished = {arXiv:2405.16444},
  url = {https://arxiv.org/abs/2405.16444}
}

@misc{teng2026infoflow,
  author = {Teng, Xin and Zhang, Canyu and Zheng, Shaoyi
    and Zhuo, Danyang and Zhou, Tianyi and Wan, Shenji},
  title = {{InfoFlow KV}: Information-Flow-Aware
    {KV} Recomputation for Long Context},
  year = {2026},
  howpublished = {arXiv:2603.05353},
  url = {https://arxiv.org/abs/2603.05353}
}

@misc{wang2025graphkv,
  author = {Wang, Haoyu and Wang, Peihao and Li, Mufei
    and Liu, Shikun and Miao, Siqi and Wang, Zhangyang
    and Li, Pan},
  title = {{Graph-KV}: Breaking Sequence via Injecting
    Structural Biases into Large Language Models},
  year = {2025},
  howpublished = {arXiv:2506.07334},
  url = {https://arxiv.org/abs/2506.07334}
}

@misc{liu2023middle,
  author = {Liu, Nelson F. and Lin, Kevin and Hewitt, John
    and Paranjape, Ashwin and Bevilacqua, Michele
    and Petroni, Fabio and Liang, Percy},
  title = {Lost in the Middle:
    How Language Models Use Long Contexts},
  year = {2023},
  howpublished = {arXiv:2307.03172},
  url = {https://arxiv.org/abs/2307.03172}
}

@misc{yu2024repetition,
  author = {Yu, Sangwon and Kim, Ik-hwan and Song, Jongyoon
    and Lee, Saehyung and Park, Junsung and Yoon, Sungroh},
  title = {Unleashing Multi-Hop Reasoning Potential
    in Large Language Models through Repetition
    of Misordered Context},
  year = {2024},
  howpublished = {arXiv:2410.07103},
  url = {https://arxiv.org/abs/2410.07103}
}

@misc{guo2025twohop,
  author = {Guo, Tianyu and Zhu, Hanlin and Zhang, Ruiqi
    and Jiao, Jiantao and Mei, Song and Jordan, Michael I.
    and Russell, Stuart},
  title = {How Do {LLMs} Perform Two-Hop Reasoning in Context?},
  year = {2025},
  howpublished = {arXiv:2502.13913},
  url = {https://arxiv.org/abs/2502.13913}
}

@inproceedings{min2019composition,
  author = {Min, Sewon and Wallace, Eric and Singh, Sameer
    and Gardner, Matt and Hajishirzi, Hannaneh
    and Zettlemoyer, Luke},
  title = {Compositional Questions Do Not Necessitate
    Multi-hop Reasoning},
  booktitle = {Proceedings of ACL},
  pages = {4249--4257},
  year = {2019},
  doi = {10.18653/v1/P19-1416},
  url = {https://aclanthology.org/P19-1416/}
}

@article{trivedi2022musique,
  author = {Trivedi, Harsh and Balasubramanian, Niranjan
    and Khot, Tushar and Sabharwal, Ashish},
  title = {{MuSiQue}: Multihop Questions
    via Single-hop Question Composition},
  journal = {Transactions of the Association
    for Computational Linguistics},
  volume = {10},
  pages = {539--554},
  year = {2022},
  doi = {10.1162/tacl_a_00475},
  url = {https://aclanthology.org/2022.tacl-1.31/}
}

@misc{qwen3technicalreport,
      title={Qwen3 Technical Report}, 
      author={{Qwen Team}},
      year={2025},
      eprint={2505.09388},
      archivePrefix={arXiv},
      primaryClass={cs.CL},
      url={https://arxiv.org/abs/2505.09388}, 
}

@misc{wei2022cot,
  author = {Wei, Jason and Wang, Xuezhi and Schuurmans, Dale
    and Bosma, Maarten and Ichter, Brian and Xia, Fei
    and Chi, Ed and Le, Quoc and Zhou, Denny},
  title = {Chain-of-Thought Prompting Elicits Reasoning
    in Large Language Models},
  year = {2022},
  howpublished = {arXiv:2201.11903},
  url = {https://arxiv.org/abs/2201.11903}
}

@inproceedings{
    ma2025blockattention,
    title={Block-Attention for Efficient Prefilling},
    author={Dongyang Ma and Yan Wang and Tian Lan},
    booktitle={The Thirteenth International Conference on Learning Representations},
    year={2025},
    url={https://openreview.net/forum?id=7zNYY1E2fq}
}

@misc{peng2026kvcacherepairbudgetgo,
      title={Where Should a KV-Cache Repair Budget Go? Distractor Structure and Contextual Compatibility in Non-Prefix Reuse}, 
      author={Yuxing Peng and Zhiqing Tang and Weijia Jia},
      year={2026},
      eprint={2611.06397},
      archivePrefix={arXiv},
      primaryClass={cs.CL},
      url={https://arxiv.org/abs/2611.06397}, 
}

@inproceedings{izacard2021fid,
  author = {Izacard, Gautier and Grave, Edouard},
  title = {Leveraging Passage Retrieval with Generative Models
           for Open Domain Question Answering},
  booktitle = {Proceedings of the 16th Conference of the European
               Chapter of the Association for Computational
               Linguistics: Main Volume},
  year = {2021},
  pages = {874--880},
  address = {Online},
  publisher = {Association for Computational Linguistics},
  doi = {10.18653/v1/2021.eacl-main.74},
  url = {https://aclanthology.org/2021.eacl-main.74/}
}

@inproceedings{trivedi2023ircot,
  author = {Trivedi, Harsh and Balasubramanian, Niranjan
            and Khot, Tushar and Sabharwal, Ashish},
  title = {Interleaving Retrieval with Chain-of-Thought Reasoning
           for Knowledge-Intensive Multi-Step Questions},
  booktitle = {Proceedings of the 61st Annual Meeting of the
               Association for Computational Linguistics
               (Volume 1: Long Papers)},
  year = {2023},
  pages = {10014--10037},
  address = {Toronto, Canada},
  publisher = {Association for Computational Linguistics},
  doi = {10.18653/v1/2023.acl-long.557},
  url = {https://aclanthology.org/2023.acl-long.557/}
}

@misc{cheng2026latent,
  author = {Jiaming Cheng and Subhransu Das and Rajiv Ramnath},
  title = {When Does Latent Communication Pay? A Causal Audit
           of Relayed {KV} Caches in Multi-Agent {LLM}s},
  year = {2026},
  eprint = {2608.04893},
  archivePrefix = {arXiv},
  primaryClass = {cs.CR},
  url = {https://arxiv.org/abs/2608.04893}
}

@article{grattafiori2024llama3herdmodels,
  title={The Llama 3 Herd of Models},
  author={Grattafiori, Aaron and Dubey, Abhimanyu and Jauhri, Abhinav and others},
  journal={arXiv preprint arXiv:2407.21783},
  year={2024}
}

@inproceedings{
    wang2026prophetkv,
    title={Prophet{KV}: User-Query-Driven Selective Recomputation for Efficient {KV} Cache Reuse in Retrieval-Augmented Generation},
    author={Shihao Wang and Jiahao Chen and Yanqi Pan and Hao Huang and Yichen Hao and Xiangyu Zou and Wen Xia and Chongyang Qiu and Wentao Zhang and PengFei Wang},
    booktitle={Forty-third International Conference on Machine Learning},
    year={2026},
    url={https://openreview.net/forum?id=SwGwmKIgUb}
}

@misc{kim2026coinragcontextualizedinformationnugget,
      title={CoinRAG: Contextualized Information Nugget KV Cache Reuse for Long-Context RAG}, 
      author={Gyuwan Kim and Cheoneum Park and Tao Yang},
      year={2026},
      eprint={2608.07458},
      archivePrefix={arXiv},
      primaryClass={cs.CL},
      url={https://arxiv.org/abs/2608.07458}, 
}

@inproceedings{yang-etal-2018-hotpotqa,
    title = "{H}otpot{QA}: A Dataset for Diverse, Explainable Multi-hop Question Answering",
    author = "Yang, Zhilin  and
      Qi, Peng  and
      Zhang, Saizheng  and
      Bengio, Yoshua  and
      Cohen, William  and
      Salakhutdinov, Ruslan  and
      Manning, Christopher D.",
    editor = "Riloff, Ellen  and
      Chiang, David  and
      Hockenmaier, Julia  and
      Tsujii, Jun{'}ichi",
    booktitle = "Proceedings of the 2018 Conference on Empirical Methods in Natural Language Processing",
    month = oct # "-" # nov,
    year = "2018",
    address = "Brussels, Belgium",
    publisher = "Association for Computational Linguistics",
    url = "https://aclanthology.org/D18-1259/",
    doi = "10.18653/v1/D18-1259",
    pages = "2369--2380"
}

@misc{baquero2026cachemergingconvergentreplicated,
      title={Cache Merging as a Convergent Replicated State for Multi-Agent Latent Reasoning}, 
      author={Carlos Baquero and Luís Brito},
      year={2026},
      eprint={2607.01308},
      archivePrefix={arXiv},
      primaryClass={cs.MA},
      url={https://arxiv.org/abs/2607.01308}, 
}

@misc{qi2026relaxkvrecomputationguidedquery,
      title={RelaxKV: Recomputation Guided by the Query with Sparse Context Attention for Efficient KV Cache Reuse}, 
      author={Ruoling Qi and Yirui Liu and Xuaner Wu and Yuxin Jin and Jian Chen and Jiayu Qin and Yin Chen and Jiawei Shao},
      year={2026},
      eprint={2609.33503},
      archivePrefix={arXiv},
      primaryClass={cs.AI},
      url={https://arxiv.org/abs/2609.33503}, 
}

@misc{pan2026cacheweavelearningrecomputeefficient,
      title={CacheWeave: Learning to Recompute for Efficient Position-Independent KV Caching}, 
      author={Yueyang Pan and Qianchao Zhu and Yuliang Liu and Teng Ma and Sanidhya Kashyap},
      year={2026},
      eprint={2611.01735},
      archivePrefix={arXiv},
      primaryClass={cs.DC},
      url={https://arxiv.org/abs/2611.01735}, 
}

\setcounter{secnumdepth}{1}
\appendix

\section{Implementation and Runtime Provenance}
\label{app:setup}

\paragraph{Runtime.} The MuSiQue Qwen3-4B run uses float16 on an NVIDIA L4 with PyTorch 2.6.0+cu124 and CUDA 12.4. The two MuSiQue 8B runs use bfloat16 on an A100-SXM4-40GB with PyTorch 2.14.1+cu130 and CUDA 13.0. All three HotpotQA runs use an A100-SXM4-80GB with PyTorch 2.14.1+cu130, with float16 for Qwen3-4B and bfloat16 for the 8B models. The position, co-location, and additional mask checks use A100 80GB hardware. All use Transformers 4.57.6, Tokenizers 0.22.2, scaled dot-product attention, batch size one, and greedy decoding. The archives record checkpoint revisions and tokenizer hashes. Deterministic algorithms and TF32 matrix multiplication are disabled.

\paragraph{Masks and matched access.}
An explicit four-dimensional additive mask is shared across heads and applied at every transformer layer during evidence prefill. Paragraph spans use offsets from unchanged block tokenization; \full{} also includes separators. Answer generation continues from the resulting cache with unrestricted access to all evidence positions.

For \matched{}, candidate paragraphs precede A's designated support and contain no recognized answer-alias or bridge mention. A hash of the layout identifier selects a window in the concatenated eligible token indices, with exactly the support paragraph's token count. The window can span portions of multiple paragraphs. Selection is fixed before generation. Positions are unmatched, and the selected states can contain relevant information from preceding paragraphs. On HotpotQA, excluding the 27 sources whose pre-window text mentions the answer or bridge in any model leaves the pooled contrasts within 0.1 points: \gold{}--\matched{} is $+7.0$ $[4.3, 9.8]$ and \matched{}--\blocked{} is $-0.2$ $[-2.1, 1.6]$ on the remaining 173.
\begin{figure}[tb]
\centering
\setlength{\unitlength}{1pt}
\small

\begin{picture}(236,194)(0,30)

\put(72,188){\framebox(100,32){
  \shortstack{\textbf{Shared prefix S}\\System + question}
}}

\put(85,188){\vector(-1,-1){26}}
\put(159,188){\vector(1,-1){26}}

\put(16,103){\framebox(86,59){}}
\put(16,143){\makebox(86,14){
  \textbf{Earlier block A}
}}
\put(16,124){\makebox(86,14){Distractors}}
\put(16,107){\makebox(86,14){Support}}

\put(142,103){\framebox(86,59){}}
\put(142,143){\makebox(86,14){
  \textbf{Later block B}
}}
\put(142,124){\makebox(86,14){Distractors}}
\put(142,107){\makebox(86,14){Support}}

\put(103,135){\vector(1,0){38}}
\put(103,141){\makebox(38,12){V}}
\put(103,117){\makebox(38,12){
  \footnotesize Encoding
}}

\put(59,103){\vector(1,-1){35}}
\put(185,103){\vector(-1,-1){35}}
\put(77,85){\makebox(90,12){
  \footnotesize Answer-time access
}}

\put(61,35){\framebox(122,33){
  \shortstack{
    \textbf{Answerer}\\
    All positions in S, A, B
  }
}}

\put(72,204){\line(-1,0){68}}
\put(4,204){\line(0,-1){153}}
\put(4,51){\vector(1,0){57}}

\end{picture}

\par\smallskip
{\footnotesize
\begin{tabular}{@{}ll@{}}
\blocked{} & No positions in A\\
\gold{}    & Supporting paragraph in A\\
\matched{} & Equal-token non-support access\\
\full{}    & All positions in A
\end{tabular}
}

\caption{Encoding-access intervention in the split layout.
V specifies the positions in A accessible to B
during encoding. The answerer retains access to every
prefix and evidence position. Arrows indicate information
availability.}
\label{fig:design}
\end{figure}

\paragraph{HotpotQA selection.}
Of 351 HotpotQA bridge questions examined in the fixed order, we exclude 93 whose answer appears in both supporting paragraphs or in neither, 40 with no supporting title mentioned by the other support, 12 for which an equal-token window cannot be constructed, and 7 used in earlier runs; one question meets two criteria, leaving 200.

\paragraph{Prompts and budgets.}
The shared prefix contains the system instruction, question, and an instruction to use relevant material while ignoring irrelevant content. Paragraphs are separated by two newlines. The answer instruction is:
\begin{quote}
Give the shortest exact answer (a name, entity, number, or yes/no)
inside \texttt{\textbackslash boxed\{YOUR\_FINAL\_ANSWER\}}.
\end{quote}
The Qwen3 non-thinking protocol prefills an empty \texttt{<think>}--\texttt{</think>} block; Llama-8B uses its native assistant prefix. Ordinary answer text can still contain reasoning. Primary answers have a 1,024-token cap; shortcut probes have a 256-token cap. Forced-thinking Qwen3 experiments prefill \texttt{<think>} and transition to answer generation at closure or budget exhaustion, with separate 1,024-token thinking and answer caps. Only the extracted answer is scored.

\paragraph{Validation and truncation.}
Input-token hashes and answer positions agree across all six primary conditions in every layout; \gold{} and \matched{} expose equally many tokens in every layout. For each 8B model, the 48 content-swap checks leave B's KV states unchanged under blocking and change them under full access. Equality between \full{} and $B1$ holds with a common attention kernel; default kernel selection can produce numerical and generation differences. Equality between the blocked mask and separate encoding at the same positions remains unestablished. 

Across 2,400 primary answers per model and dataset, answer-cap hits total 8, 0, and 103 on MuSiQue and 5, 0, and 52 on HotpotQA for Qwen3-4B, Qwen3-8B, and Llama-8B, respectively. The co-location follow-up has 0 and 31 hits out of 800 answers for the two 8B models. All capped responses remain in the reported scores. Excluding sources with a capped answer in either condition leaves the Llama-8B HotpotQA \gold{}--\blocked{}, \gold{}--\matched{}, and \full{}--\blocked{} contrasts within 0.8 points of the full-cohort values, with intervals still above zero.

\section{Additional Masking and Reasoning Results}
\label{app:tables}

\paragraph{Scores by condition and multi-seed means.}
Table~\ref{tab:main} gives EM and F1 for every condition at data-layout seed 0, and Table~\ref{tab:multiseed} gives mean differences across data-layout seeds.

\begin{table}[tb]
\centering
\footnotesize
\begin{tabular}{@{}lrrr@{}}
\toprule
Condition & Qwen3-4B & Qwen3-8B & Llama-8B\\
\midrule
\multicolumn{4}{@{}l}{MuSiQue}\\
\blocked{} & 26.8 / 33.5 & 34.0 / 42.3 & 31.0 / 38.5\\
\matched{} & 25.2 / 32.0 & 34.0 / 42.7 & 29.5 / 37.1\\
\gold{} & 34.8 / 43.3 & 40.0 / 49.8 & 36.0 / 44.2\\
\full{} & 31.8 / 39.7 & 39.8 / 48.8 & 35.5 / 42.8\\
$B1$ & 31.8 / 39.8 & 40.2 / 49.5 & 35.2 / 43.4\\
$K0$ & 29.0 / 35.5 & 34.5 / 42.9 & 30.5 / 37.9\\
\midrule
\multicolumn{4}{@{}l}{HotpotQA}\\
\blocked{} & 37.0 / 49.8 & 43.5 / 59.3 & 35.2 / 50.0\\
\matched{} & 36.8 / 49.8 & 43.2 / 60.2 & 34.2 / 48.2\\
\gold{} & 44.2 / 58.5 & 48.2 / 65.9 & 41.0 / 54.9\\
\full{} & 44.0 / 58.0 & 47.5 / 64.5 & 43.0 / 56.8\\
$B1$ & 43.8 / 57.8 & 47.0 / 64.1 & 41.8 / 55.4\\
$K0$ & 37.8 / 50.9 & 43.8 / 59.5 & 35.2 / 48.4\\
\bottomrule
\end{tabular}
\caption{EM / F1 (0--100) at data-layout seed 0, averaging $AB$/$BA$ within 200 source questions per dataset. All conditions retain every supplied paragraph. $B1$ is native joint encoding; $K0$ uses stitched caches.}
\label{tab:main}
\end{table}

\begin{table}[tb]
\centering
\footnotesize
\begin{tabular}{@{}lrrrr@{}}
\toprule
Dataset & Seeds & $G{-}B$ & $G{-}M$ & $G{-}F$\\
\midrule
\multicolumn{5}{@{}l}{Qwen3-8B}\\
MuSiQue & 5 & $+7.2$ & $+7.1$ & $+0.6$\\
HotpotQA & 5 & $+6.4$ & $+6.0$ & $+0.9$\\
\midrule
\multicolumn{5}{@{}l}{Llama-8B}\\
MuSiQue & 3 & $+5.8$ & $+6.3$ & $+0.5$\\
HotpotQA & 3 & $+5.3$ & $+6.2$ & $-1.1$\\
\bottomrule
\end{tabular}
\caption{Mean F1 differences across data-layout seeds, including seed 0. $G$, $B$, $M$, and $F$ denote \gold{}, \blocked{}, \matched{}, and \full{}. Each seed reuses the same 200 questions per dataset; these are descriptive summaries without confidence intervals across seeds.}
\label{tab:multiseed}
\end{table}

\paragraph{Position and co-location protocols.}
The position follow-up moves A's support to its first paragraph without changing B. The matched control selects equally many non-support tokens near the end of $A$. These later states already had access to the support and can therefore carry supporting information. Only first and last support positions are tested; absolute position and distance are coupled.

 $AA$/$BB$ layouts place hop 1 followed by hop 2 in the final two paragraph slots of one worker, exchanging a distractor with the other worker. All 20 paragraphs and the 10/10 allocation are retained. Each 8B model evaluates 200 sources in both layouts under \blocked{} and \full{}. \full{}--\blocked{} is $+0.5$ F1 $[-2.7,3.7]$ on Qwen3-8B and $-2.1$ $[-5.9,1.7]$ on Llama-8B, averaging  $AA$/$BB$. The corresponding split-minus-together interactions are $+6.0$ $[1.1,10.9]$ and $+6.4 $ $[1.2,11.6]$.

\paragraph{Earlier screened masking study.}
An earlier study used the first 100 two-hop MuSiQue sources and retained 87/79 for Qwen3-4B\slash Qwen3-8B after closed-book and both single-support probes failed EM. These sources are disjoint from the primary 200-source cohort. Runs used float16 at 4B, bfloat16 at 8B, and a 256-token answer cap. Table~\ref{tab:appendix-masking} preserves its all-but-support control: \nogold{} exposes every A position except its support. \dist{} exposes one deterministically selected non-support paragraph; these earlier controls were not token-count matched.

\begin{table}[tb]
\centering
\small
\begin{tabular}{@{}lrrrr@{}}
\toprule
 & \multicolumn{2}{c}{Qwen3-4B}
 & \multicolumn{2}{c}{Qwen3-8B}\\
\cmidrule(lr){2-3}\cmidrule(lr){4-5}
Condition & EM & F1 & EM & F1\\
\midrule
\blocked{} & 13.2 & 22.8 & 14.6 & 26.4\\
\dist{}    &  9.2 & 20.1 & 13.9 & 26.4\\
\nogold{}  &  5.7 & 19.8 & 16.5 & 29.6\\
\gold{}    & 24.7 & 36.4 & 38.6 & 49.9\\
\full{}    & 24.1 & 36.1 & 29.1 & 41.5\\
\bottomrule
\end{tabular}
\caption{Earlier screened MuSiQue results, EM and F1 (0--100),
averaging $AB$/$BA$ within 87/79 sources at 4B/8B. These are separate
cohorts and protocols from Table~\ref{tab:main}.}
\label{tab:appendix-masking}
\end{table}

\paragraph{Synthetic reasoning follow-up.}
On 25 new two-fact toy chains evaluated with Qwen3-8B, the person-to-city and city-to-mayor supports are always in different workers. A look-alike Cities distractor is placed in the same worker as the supporting Cities paragraph or in the other worker. Crossing this placement with $AB$/$BA$, identical/distinct Cities headings, two masks, and two decoding protocols yields 800 answers (Table~\ref{tab:appendix-toy}). Under \blocked{} with forced thinking and the Cities distractor in the other worker, 60/100 responses exhaust the thinking budget: 59 are wrong, while all 40 uncapped responses are correct. In that cell, forced thinking turns 50 correct answers into incorrect ones and 2 incorrect answers into correct ones. Additionally, 39 are correct and 9 incorrect under both protocols. All 50 lost answers exhaust the thinking budget and end in an abstention such as ``Unknown'' or ``let me check again''. Across the four forced-thinking cells of Table~\ref{tab:appendix-toy}, 77 of 400 responses exhaust the budget; 71 of them contain ``let me check again'' four or more times (case-insensitive), against none of the 323 that finish. These counts are descriptive: they do not show why forced thinking lowers accuracy here, or what a longer budget would recover.

\begin{table}[tb]
\centering
\small
\begin{tabular}{@{}llrr@{}}
\toprule
Cities distractor & Mask & Off & Forced\\
\midrule
Other worker & \blocked{} & 89  & 41\\
Other worker & \gold{}    & 100 & 96\\
Same worker  & \blocked{} & 97  & 97\\
Same worker  & \gold{}    & 100 & 100\\
\bottomrule
\end{tabular}
\caption{Qwen3-8B toy EM (\%), 25 chains and 100 layouts per cell,
averaging hop order and heading variants. Distractor placement is
relative to the worker holding the supporting Cities paragraph.
The two supporting facts always remain split across workers.}
\label{tab:appendix-toy}
\end{table}

\paragraph{Earlier MuSiQue reasoning study.}
A separate 25-source Qwen3-8B study from the earlier source pool evaluated $AB$/$BA$ with thinking off or forced. F1 changed from 21.7 to 47.6 under \blocked{}, 53.8 to 57.2 under \gold{}, and 43.2 to 57.2 under \full{}. Forced thinking reached its 1,024-token budget in 22/50, 18/50, and 19/50 responses respectively. This shows recovery on that sample under the tested budget. Forced thinking has not been evaluated on either primary cohort.

\section{Screening Sensitivity}
\label{app:screening}
The primary cohorts were selected independently of model correctness. For sensitivity analysis, we retain sources whose closed-book and both single-support probes all score F1 below 0.5 without truncation. Single-support probes retain all distractors and use the $AB$ layout. Table~\ref{tab:appendix-screening} reports results on these model-specific subsets, using the primary source-bootstrap procedure. Probe failure does not establish that both supports are necessary. These overlapping subsets are sensitivity analyses, not independent replications.

\begin{table}[tb]
\centering
\footnotesize
\begin{tabular}{@{}lrcc@{}}
\toprule
Model & n & \gold{}--\blocked{} & \gold{}--\matched{}\\
\midrule
\multicolumn{4}{@{}l}{MuSiQue}\\
Qwen3-4B & 127 & $+10.1$ & $+12.2$\\
 & & $[5.8,14.5]$ & $[7.6,17.2]$\\
Qwen3-8B & 102 & $+11.5$ & $+10.1$\\
 & & $[5.8,17.3]$ & $[4.0,16.2]$\\
Llama-8B & 77 & $+9.6$ & $+10.8$\\
 & & $[3.8,15.4]$ & $[5.7,16.5]$\\
\midrule
\multicolumn{4}{@{}l}{HotpotQA}\\
Qwen3-4B & 86 & $+9.2$ & $+9.7$\\
 & & $[4.3,14.5]$ & $[4.9,14.8]$\\
Qwen3-8B & 76 & $+7.3$ & $+5.3$\\
 & & $[0.2,14.7]$ & $[-1.6,12.1]$\\
Llama-8B & 64 & $+6.6$ & $+8.0$\\
 & & $[-0.4,13.8]$ & $[0.8,15.5]$\\
\bottomrule
\end{tabular}
\caption{Paired F1 differences with 95\% intervals on stricter probe subsets of each primary cohort. n counts source questions.}
\label{tab:appendix-screening}
\end{table}
\section{Joint Encoding, Stitched Caches, and Text Reports}
\label{app:thinking4b}

An earlier Qwen3-4B experiment compares native joint encoding ($B1$), stitched caches ($K0$), and generated worker reports ($T1$). $T1$ has each worker write a report from its documents, which the answerer reads jointly. $B1$--$K0$ changes both encoding access and cache-position handling.

The study uses the first 50 sources per task group: two-fact toys, three-fact toys, and two-hop MuSiQue. It evaluates $AB$/$BA$ and an additional MuSiQue $BB$ placement, yielding 150 source identifiers, 350 layouts, and 2,100 responses. It uses no correctness-based probe screen and is separate from the primary 200-source cohort. The three-fact toy adds a mayor-to-birthplace fact to the People paragraph containing the person-to-city fact, so the chain returns to that worker.

All conditions use Qwen3-4B in float16 on an NVIDIA T4, greedy decoding, and data seed 0. Forced thinking and final answers have separate 1,024-token budgets; worker reports also have a 1,024-token cap. In $T1$, only the answerer's thinking protocol changes: all 350 worker-report pairs are identical across decoding conditions.

\begin{table}[t]
\centering
\small
\begin{tabular}{@{}lrrr@{}}
\toprule 
Metric & $B1$ & $K0$ & $T1$\\
\midrule
\multicolumn{4}{@{}l}{Two-fact toy / $AB$+ $BA$}\\
EM & 100.0 / 100.0 & 29.0 / 74.0 & 45.0 / 90.0\\
\midrule
\multicolumn{4}{@{}l}{Three-fact toy / $AB$+ $BA$}\\
EM & 100.0 / 100.0 & 21.0 / 60.0 & 51.0 / 74.0\\
\midrule
\multicolumn{4}{@{}l}{MuSiQue / $AB$+ $BA$}\\
EM & 35.0 / 46.0 & 17.0 / 36.0 & 22.0 / 21.0\\
F1 & 49.6 / 62.1 & 29.5 / 51.8 & 37.4 / 37.0\\
\midrule
\multicolumn{4}{@{}l}{MuSiQue / $BB$}\\
EM & 26.0 / 42.0 & 34.0 / 50.0 & 46.0 / 48.0\\
F1 & 40.7 / 57.4 & 46.7 / 67.5 & 57.7 / 60.2\\
\bottomrule
\end{tabular}
\caption{Earlier Qwen3-4B experiment. Cells report thinking off / forced thinking (0--100). $B1$: joint encoding; $K0$: stitched caches; $T1$: text reports. Each task has 50 sources; $AB$ + $BA$ includes 100 layouts and $BB$ includes 50. MuSiQue placements reuse the same sources. Toy EM and F1 coincide.}
\label{tab:appendix-thinking4b}
\end{table}

\paragraph{Paired effects and generation budgets.} For this earlier experiment, intervals use 10,000 source-bootstrap resamples with seed 20261116, averaging $AB$/$BA$ within source. Forced thinking improves $K0$ EM by 45 points $[33,56]$ on two-fact toys, 39 $[28,49]$ on three-fact toys, and 19 $[9,30]$ on split MuSiQue. Under forced thinking, $B1$ retains advantages over $K0$ of 26 $[19,33]$, 40 $[31,49]$, and 10 $[2,18]$ respectively. $T1$ changes by $-1$ point $[-8,6]$ on split MuSiQue. The $BB$ results show that joint encoding is not uniformly best; that placement also changes positions and distractor allocation.

Mean thinking-token counts are 400/651/690 for $B1$/$K0$/$T1$, with 41/127/145 thinking-cap hits out of 350 responses each. Four $K0$ final answers are truncated and remain in the scores; no worker report is truncated. Shared maximum budgets therefore do not imply equal realized computation. These results show substantial recovery in this experiment, while leaving unrestricted-budget performance unresolved.

\section{Exploratory Analysis of \gold{} and \full{} Disagreements}
\label{app:error-analysis}

Table~\ref{tab:gold-full-disagreements} summarizes exact-match outcomes for all 400 primary MuSiQue layouts per model, using the original scorer and reference aliases and retaining capped answers. Disagreements occur in both directions. These are descriptive layout counts: each source contributes two paired $AB$/$BA$ layouts, so the 400 layouts are not independent observations.

\begin{table}[tb]
\centering
\small
\begin{tabular}{@{}lrrrr@{}}
\toprule
Model & Gold only & Full only & Both & Neither\\
\midrule
Qwen3-4B & 25 & 13 & 114 & 248\\
Qwen3-8B & 23 & 22 & 137 & 218\\
Llama-8B & 29 & 27 & 115 & 229\\
\bottomrule
\end{tabular}
\caption{Exact-match outcomes for \gold{} and \full{}. ``Gold only'' means that only \gold{} matches a reference alias; ``Full only'' is the converse. ``Both'' and ``Neither'' indicate matching under both conditions or under neither. Each row covers 200 MuSiQue sources evaluated in $AB$/$BA$ layouts.}
\label{tab:gold-full-disagreements}
\end{table}

\paragraph{Illustrative saved outputs.}
The following Qwen3-4B cases were selected retrospectively to illustrate different answer disagreements. They do not estimate the prevalence of underlying error mechanisms.

\paragraph{Distractor-associated answer.}
For source \texttt{2hop\_\_61714\_64633} in $AB$ order, the question asks which episode of \emph{The Office} depicts Dwight saving Pam's husband from Roy. The supporting paragraphs identify Jim as Pam's husband and describe the incident in ``The Negotiation.'' \gold{} returns ``The Negotiation,'' whereas \full{} returns ``The Farm,'' an episode named in two non-supporting paragraphs in $A$. This is compatible with distractor interference but does not establish that mechanism: those paragraphs remain available to the answerer in both conditions.

\paragraph{Intermediate versus final answer.}
The reverse outcome occurs for source \texttt{2hop\_\_157285\_556157} in  $BA$ order. The supplied evidence connects Mtetengwe to the Mzingwane River and Mzingwane to the Limpopo River. \gold{} returns the intermediate entity, ``Mzingwane River,'' while \full{} returns the reference answer, ``Limpopo.'' This illustrates a case in which full access yields the correct final answer.

\paragraph{Reference-alias mismatch.}
For source \texttt{2hop\_\_133116\_533882} in $AB$ order, \gold{} returns ``Martin Marietta'' and \full{} returns ``Martin Marietta Corporation.'' The supporting paragraph identifies the same company, but the saved reference contains only ``Martin Marietta.'' Exact match therefore accepts only \gold{}, while token F1 assigns 100 and 80, respectively, on the 0--100 scale. This disagreement reflects reference coverage rather than a clearly incorrect entity prediction.

These examples illustrate limitations of interpreting aggregate score differences. They neither identify an internal attention mechanism nor establish a general advantage for selective access.


\end{document}